# Energy localization and excess fluctuations from long-range interactions in equilibrium molecular dynamics

Ralph V. Chamberlin

Department of Physics, Arizona State University, Tempe, AZ 85287-1504 USA

Vladimiro Mujica

School of Molecular Science, Arizona State University, Tempe, AZ 85287-1604 USA

Sergei Izvekov and James P. Larentzos

US Army Research Laboratory, Aberdeen Proving Ground, MD 21005 USA

**Abstract**

Molecular dynamics (MD) simulations of standard systems of interacting particles ("atoms") give excellent agreement with the equipartition theorem for the average energy, but we find that these simulations exhibit finite-size effects in the dynamics that cause local fluctuations in energy to deviate significantly from the analogous energy fluctuation relation (EFR). We have made a detailed analysis of Lennard-Jones atoms to track the origin of such unphysical fluctuations, which must be corrected for an appropriate description of the statistical mechanics of small systems, especially at low temperatures ($T$). Similar behavior is found in a model of nitromethane at higher $T$. The main conclusion of our study is that systems separated into nanometer-sized "blocks" inside much larger simulations exhibit excess fluctuations in potential energy ($pe$) that diverge inversely proportional to $T$ in a manner that is strongly dependent on the range of interaction. Specifically, at low $T$ with long-range interactions $pe$ fluctuations exceed the EFR by at least an order of magnitude, dropping abruptly to below the EFR when interactions include only 1st-neighbor atoms. Thus, excess $pe$ fluctuations cannot be due to simple surface effects from the robustly harmonic 1st-neighbor interactions, nor from any details in the simulations or analysis. A simplistic model that includes 2nd-neighbor interactions matches the excess fluctuations, but only if the 2nd-neighbor energy is not included in Boltzmann's factor, attributable to anharmonic effects and energy localization. Empirically, adding 2nd-neighbor interactions considerably increases the width of the $pe$ distribution, suggesting that Anderson localization may play a role. Characterizing energy correlations as a function of time and distance reveals that excess $pe$ fluctuations in each block coincide with negative $pe$ correlations between neighboring blocks, whereas reduced $pe$ fluctuations coincide with positive $pe$ correlations. Indeed, anomalous $pe$ fluctuations in small systems can be quantified using a net local energy in Boltzmann's factor that includes $pe$ from the surrounding shell of similarly small systems, or equivalently an effective local temperature. Our results and analysis elucidate the source of non-Boltzmann fluctuations, and the need to include mesoscopic thermal effects from the local environment for a consistent theoretical description of the equilibrium fluctuations in MD simulations of standard models with long-range interactions.

**Keywords:** statistical physics; MD simulations; fluctuations; nanoscale dynamics; specific heat; crystals

# I. Introduction

Molecular dynamics (MD) simulations provide a powerful tool for studying the behavior of atoms and molecules in many models of practical importance, but they also provide idealized systems for fundamental investigations connecting classical and statistical mechanics [1-5]. In fact, the first MD simulations using a digital computer tested whether point-like particles obeying classical mechanics would reach thermal equilibrium if connected by anharmonic springs. To the surprise of Enrico Fermi and his team, these simulations showed recurrent energy cycling with no indication of the stable equipartition of energy expected for classical systems in the canonical ensemble. Now it is known that chaotic motion of interacting particles ("atoms") yields thermal-equilibrium average behavior as an emergent phenomenon of large systems, but questions remain regarding the local dynamics [6-10]. Here we use MD simulations of diverse models to test for canonical-ensemble behavior in the average local energies, and their fluctuations. The average energies show excellent agreement with the expected equipartition theorem, but their fluctuations often deviate from the usual energy fluctuation relation (EFR), sometimes by more than an order of magnitude. Key insight comes from energy correlations as a function of time and distance showing that deviations from this EFR can be attributed to energy that is transiently localized during equilibrium fluctuations, thereby altering the net energy that reaches distant parts of the simulation volume serving as the heat reservoir needed for the emergence of a well-defined Boltzmann's factor. Another result is a connection between energy fluctuations and atom density due to the attractive interaction between $2^{nd}$-neighbor atoms. This interpretation is confirmed by reducing the interaction range until there are only repulsive forces between $1^{st}$-neighbor atoms, which is the only way we have found to obtain dynamics that fully agrees with the EFR.

Because conservation of total energy is intrinsic to MD simulations based solely on Newton's laws, fluctuations in the entire simulation volume come only from transfer between kinetic energy ($KE$) and potential energy ($PE$). (Lower-case acronyms will be reserved for local energies inside the simulation where we use energy per particle.) Theoretical expressions have been derived for the expected canonical-ensemble specific heat from $KE$ fluctuations in microcanonical simulations [11-13]. However, using these expressions for MD simulations at low temperatures ($T$) yield specific heats that are much less than expected from the EFR, as reported previously [14], which can be attributed to excess $KE$ fluctuations from the excess $pe$ fluctuations that we study here. Anomalously large fluctuations in energy, which increase linearly proportional to the number of particles in the system ($n$), are expected in the theory of small-system thermodynamics ("nanothermodynamics") [15,16] for the fully-open nanocanonical ensemble due to the unrestricted size of the system [17]. Here we avoid this source of excess fluctuations by studying systems with fixed volume and constrained $n$.

To focus on local energy fluctuations we subdivide large simulations into smaller volumes called "blocks." Similar subdivisions have long been used for theoretical investigations of fluctuations [18], and for deriving the canonical and grand-canonical ensembles [19,20]. Early applications to MD simulations of the Lennard-Jones (L-J) model revealed anomalous size-dependent fluctuations in energy and atom density [21,22], but their main focus was to extrapolate to bulk behavior, whereas here we are interested in local effects. Studies of the L-J model using the generalized Langevin equation for a central system and its local bath agree with Maxwell-Boltzmann behavior and expected fluctuations in *ke* [23], but fluctuations in *pe* were not reported. Investigations of small systems at low *T* have found anomalous fluctuations in both *ke* and *pe* [24-26], with various explanations proposed including broken ergodicity, a need to rescale *T*, and transient normal-modes that become quantum-like at low *T*. Here we avoid these issues by studying thermal-equilibrium fluctuations about the ground-state energy of small systems ($n$=1 to 6912 atoms) inside much larger simulations that are in the thermodynamic limit, as shown by negligible dependence on the total size of the simulation ($N$=55,296-442,368 atoms). A reduction in *pe* fluctuations above the critical point for $n$<1000 has been found in Monte Carlo (MC) simulations of the Ising model for binary "spins" on a lattice [27]. Similar behavior has been attributed to a local thermal bath using nanothermodynamics [28], with reduced fluctuations because the spins are on a fixed lattice. Adding a nonlinear correction to Boltzmann's factor increases the fluctuations, making them consistent with entropy that is extensive and additive. Furthermore, this correction significantly improves agreement between MC simulations of standard models and the measured response of many materials, including critical fluids and ferromagnets [29,30]. Moreover, a related correction provides a common foundation for $1/f$ noise found at low frequencies in most substances [31-33].

Over the past few decades, several fluctuation theorems have been derived for non-equilibrium dynamics [34-40]. Although these theorems should also apply to equilibrium fluctuations, they usually rely on Boltzmann's factor at some point in the derivation, which requires weak but immediate thermal contact to an effectively infinite heat reservoir [41-44], unlike the behavior found here for the local dynamics of MD simulations. All models we study show significant localization of energy in space, and in time, especially at low *T* where neighboring atoms move about effectively harmonic *pe* minima. Thus, further study will be needed to decide if there is any connection to the localized modes in highly-anharmonic lattices [45-47]. Similarly, because we find that vibrational modes in crystals are maximally localized at lowest *T*, where disorder should be minimized, further study will also be needed to clarify any connection to Anderson localization that generally involves highly-disordered materials [48,49].

Excess energy fluctuations in MD simulations of model crystals may seem similar to measurements of high-purity single crystals at *T*<1 K, where heat capacities can exceed Debye's theory by an order of magnitude or more [50,51], but various details indicate that there is no clear connection. One detail is that measured heat capacities increase when impurities are added, so that residual deviations from Debye's theory could come from unavoidable imperfections in even the best crystals available in the laboratory, while simulations can be done on ideal crystals. More importantly, the experiments measure heat capacity, not energy fluctuations; whereas MD simulations exhibit fundamental deviations primarily from the EFR. Measurements of time-dependent specific heat [52-54] and nonresonant spectral hole burning [55-59] have shown that energy is persistently localized on time-scales of the primary response (e.g. 100 μs to 100 s) in most types of materials, including liquids, glasses, polymers, and crystals. Other techniques [60-65] have established that this energy localization occurs in correlated "regions," as defined by having dynamics that is uncorrelated with neighboring regions, which often have length scales of nanometers (e.g. 10 molecules to 390 monomer units [66]) but need not be cuboid in shape. Although superficially similar to the energy localization we find in MD simulations, various features indicate that there is no clear connection. One feature is that energy localization in measurements persists for times that are several orders of magnitude longer than in the MD simulations, but this may be due to practical limits on the speed of the measurements and the simplicity of the simulated systems. Another feature is that energy localization is most conspicuous in disordered systems; but as in the measurements of low-$T$ heat capacity, increasing disorder may merely increase the magnitude of the effect, with significant energy localization occurring in even the best available crystals. More importantly, various experimental techniques have shown that the localization is associated with abrupt decorrelation across sharp boundaries between neighboring regions in the measurements [67-69], unlike the strong correlations between neighboring blocks in MD simulations. In fact, we suggest that models with interactions that facilitate this decorrelation can improve the agreement between MD simulations, theory, and measured thermal and dynamic properties of real materials, as has been done for MC simulations [28-33].

      The main conclusion of our study is that excess *pe* fluctuations occur for all types of small systems inside large simulations, but only if there are long-range interactions. The small systems include individual atoms, and fixed-volume blocks containing 8-6912 atoms, which fluctuate about equilibrium inside much larger simulations that are in the thermodynamic limit, using either pure Newtonian dynamics or a Nosé-Hoover thermostat. Excess *pe* fluctuations are correlated with atom density, and depend strongly on the interaction cutoff radius. Specifically, *pe* fluctuations in small systems with long-range interactions exceed the EFR by at least an order of magnitude at low *T*, dropping abruptly to below the EFR when interactions include only 1$^{st}$-neighbor atoms. Thus, excess *pe* fluctuations cannot come from simple surface effects involving robustly harmonic 1$^{st}$-neighbor interactions, nor from any details in the simulations or analysis. Instead, the mechanism involves the intrinsically anharmonic interactions between 2$^{nd}$-neighbor atoms, attributable to the local energy reduction when these atoms fluctuate towards the *pe* minimum at the 1$^{st}$-neighbor distance. Additional evidence comes from high-frequency normal modes that are found to

dominate the excess *pe* fluctuations, and connect quantitatively to specific zone-boundary vibrations that couple primarily to 2nd-neighbor interactions.

## II. Background

Standard statistical mechanics is based on Boltzmann's factor, $e^{-E/kT}$, where $k$ is Boltzmann's constant. Boltzmann's factor yields the normalized probability that a given system has energy $E$, $p_E = e^{-E/kT}/Z$, where $Z$ is the partition function, $Z = \int e^{-E'/kT} g_{E'} dE'$. Here, the density of states often increases as a power-law, $g_E \propto E^\gamma$, with the exponent $\gamma = nD/2 - 1$ for an ideal gas of $n$ atoms moving in $D$ dimensions [19]. The average energy becomes

$$<E> = \int E' e^{-E'/kT} g_{E'} dE' / Z. \qquad \text{Eq. (1)}$$

According to Feynman [70]: "This fundamental law is the summit of statistical mechanics…" However, Feynman then lists some of the assumptions needed for Eq. (1). "If a system is very weakly coupled to a heat bath at a given 'temperature,' if the coupling is indefinite or not known precisely, if the coupling has been on for a long time, and if all the 'fast' things have happened and all the 'slow' things not, the system is said to be in thermal equilibrium." Thus, three assumptions that may apply to long-time average properties in MD simulations, but not to short-time dynamics, are that the transfer of energy from a local fluctuation to the large heat reservoir is too slow to govern the fast dynamics, so that the dynamics is dominated by a small local bath that interacts strongly with the system, and is altered by the fluctuations so as not to have a unique 'temperature.'

Fluctuation relations connect the internal fluctuations of a thermal quantity to its change with respect to an external parameter [71,72]. Early examples include Einstein's explanation for Brownian motion, which facilitated the first quantitative evidence for atoms [73], and Nyquist's explanation for electron fluctuations that yield Johnson noise [74]. The EFR can be derived from Eq. (1) by taking the derivative: $\frac{d<E>}{d(kT)} = \frac{\int (E')^2 e^{-E'/kT} g_{E'} dE'}{Z(kT)^2} - \left[\frac{\int E' e^{-E'/kT} g_{E'} dE'}{ZkT}\right]^2 = \frac{<E^2> - <E>^2}{(kT)^2}$, yielding:

$$\frac{d<E>}{d(kT)} = \frac{<(\Delta E)^2>}{(kT)^2}. \qquad \text{Eq. (2)}$$

Because Eq. (2) is obtained from Eq. (1) using only the product rule of calculus, this EFR is an identity that should apply to any system that exhibits canonical-ensemble dynamics.

Another simplifying feature of the canonical ensemble is that contributions from the kinetic and potential energies are independent. Because the energies add linearly, the left side of Eq, (1) becomes: *<KE+PE> = <KE> + <PE>*. Again taking the derivative of these averages, assuming only the product rule of calculus, Eq. (2) also applies separately to kinetic and potential energies. More generally, from the right side of Eq. (2), *<(ΔKE+ΔPE)²> = <(ΔKE)²> + 2<ΔKEΔPE> + <(ΔPE)²>*, so that the EFR applies separately to *KE* and *PE* whenever they fluctuate independently, intrinsic to canonical-ensemble behavior. In fact we will show that in small blocks at low *T*, $\langle\Delta KE \Delta PE\rangle \approx 0$ only if there are long-range interactions, yet this is also where the simulations show largest deviations from the EFR.

The equipartition theorem predicts that average energies have a contribution equal to $\frac{1}{2}kT$ for each classical degree of freedom. The theorem applies to any variable (*s*) that occurs quadratically in the energy, $E = E_s + E_{(s)}$, where $E_s = \frac{1}{2}Ks^2$ and $E_{(s)}$ is the energy from all variables other than *s*. Here *K* is a constant, e.g. the mass of the atom if *s* is a velocity, or a spring constant if *s* is a displacement from equilibrium. This theorem can be obtained from Eq. (1) by integrating over all energies using the usual density of states, yielding $<E_s> = \frac{\int_0^\infty (E_{s'})^{\gamma+1} e^{-E_{s'}/kT} dE_{s'}}{\int_0^\infty (E_{s'})^{\gamma} e^{-E_{s'}/kT} dE_{s'}} = kT(\gamma+1) = \frac{1}{2}nDkT$. Thus at low *T*, where neighboring atoms move in *D*=3 dimensions about a potential-energy minimum that is quadratic, Eq. (2) can be written as:

$$\frac{d<KE>}{d(kT)} + \frac{d<PE>}{d(kT)} = n\frac{3}{2} + n\frac{3}{2} = \frac{<(\Delta KE)^2>}{(kT)^2} + \frac{<(\Delta PE)^2>}{(kT)^2}. \qquad \text{Eq. (3)}$$

Theoretically, Eq. (3) is comprised of standard results found in most textbooks on statistical mechanics; empirically, it is the law of Dulong and Petit for measured heat capacities of solids at room temperature. Similar expressions have been derived for energy fluctuations of small systems in the canonical ensemble [75], using nanothermodynamics [17], and an expanded version of thermodynamics [76]. We find that the left-side of Eq. (3) is obeyed by MD simulations, but the right side is not.

## III. Simulations

All models were simulated using the Large-scale Atomic/Molecular Massively Parallel Simulator (LAMMPS) [77], with default settings for double precision math and Verlet integration. Here we focus on the standard 12-6 L-J model, with a potential energy that depends on the reduced (dimensionless) distance (*r*) between interacting atoms, $PE = 4\varepsilon\left[\left(\frac{1}{r}\right)^{12} - \left(\frac{1}{r}\right)^{6}\right]$, out to a cutoff radius $r < r_c$. For argon the energy scale is set by $\varepsilon/k$=119.8 K while *r*=1 (where *PE*=0) corresponds to 0.3405 nm [78]. A shifted force added to this model ("smooth/linear" option) ensures that both the potential and its derivative go to zero at $r=r_c$. The equilibrium distance between 1st- and 2nd-neighbor atoms is $r_0 \approx 2^{1/6} \approx 1.1225$ and $r_2 \approx 2^{4/6} \approx 1.5874$, respectively. In LAMMPS, potential energy per atom is found by splitting the energy equally between the interacting pair, *pe* = *PE*/2. This is the simplest way to partition energy between interacting atoms. Furthermore, it yields behavior consistent with energy fluctuations from individual atoms, and from blocks of varying size. The equilibrium structure of the solid phase is a face-centered-cubic (fcc) lattice. Here we focus on two simulation sizes, a cubic volume with sides of length *L*=24 yielding $24^3$=13,824 unit cells containing *N*=55,296 atoms, and *L*=48 yielding $48^3$=110,592 unit cells containing *N*=442,368 atoms. Periodic boundary conditions were used for the outside surfaces in all directions. Many simulations utilized a relatively long cutoff $r_c$=6.0, but crucial information comes from studying how *pe* fluctuations decrease with decreasing cutoff, at least until there are interactions between only 1st-neighbor atoms, $r_c<r_2\approx1.6$. In contrast, because the potential is shifted to equal zero at the cutoff, the average energy per atom at $kT/\varepsilon$=0.001 increases from $\langle pe\rangle/\varepsilon = -8.461$ for $r_c$=6.0 to $-1.535$ for $r_c$=1.5. Nevertheless, at such low *T* the potential well is still much deeper than the fluctuations, so that even for a single unit cell, *pe* varies by less than 0.4 % of its magnitude until $r_c<1.4$.

We investigate local thermal fluctuations by subdividing the entire simulation volume into smaller cube-shaped "blocks" (chunk/atom compute in LAMMPS), as sketched in Fig. 1. Specific blocks described here have sides of length $\ell$=1, 2, 3, 4, 6, 8, and 12 unit cells, yielding the number of atoms in equilibrium for the fcc lattice of $\langle n\rangle$=4, 32, 108, 256, 864, 2048, and 6912, respectively. (For simulations, angled brackets denote ensemble averages over all equivalent blocks and time averages over the entire production run.) Block positions were usually chosen with walls aligned between crystalline planes, so that most blocks strictly obey the canonical ensemble requirement of fixed *n*. In other simulations the block walls were aligned with the static crystalline planes, facilitating the study of fluctuations in *n* from even the tiniest motion of the atoms. In this case, canonical ensemble behavior is extracted by averaging energies from the subset of blocks that have essentially the correct number of atoms, e.g. $n=\langle n\rangle\pm0.5$, yielding fluctuations in *pe* and *ke* that are consistent with fixed *n*. When blocks are aligned with the crystalline planes, just filling the entire simulation volume, the number of each type ranges between 13,824 blocks for $\ell$=1 in the $L$=24 lattice (or $\ell$=2 in the $L$=48 lattice) to 27 blocks for $\ell$=8 in the $L$=24 lattice (or 64 blocks for $\ell$=12 when $L$=48 lattice). Relatively smooth time-dependent behavior is obtained by averaging over intervals of 10 to 100 time-steps. Several quantities for each block are averaged and recorded after each time interval, including the number of atoms *n*, potential energy per atom *pe*, and local temperature from the kinetic energy per atom $kT=2ke/3$. The blocks are purely mathematical, having no influence on the dynamics, so that atoms and energy can transfer freely between blocks, unaffected by block definition. Due to sample symmetry and periodic boundary conditions, each block represents an equivalent central "system," with neighboring blocks providing a self-consistent bath of energy and atoms, thereby avoiding any issues from artificial thermal baths that may alter intrinsic behavior. Theoretically, such a fixed volume with imaginary walls in a large reservoir of heat and particles is the basis for deriving the grand-canonical ensemble [19], with canonical-ensemble behavior extracted by selecting the subset of blocks with fixed *n*. Similar fluctuations in energy are found for simulations in the canonical ensemble using the Nosé-Hoover thermostat, and for individual atoms dispersed throughout the system, establishing that the behavior does not depend on block subdivision or energy sharing between atoms or blocks.

Nitromethane (NM) was simulated using interactions and input parameters that are chosen for good agreement with bulk experimental properties as a function of temperature and pressure [79]. The intramolecular potentials come from a superposition of bond-stretching, bond-bending, and torsional-angle terms. The intermolecular potentials include the Buckingham 6-exp form, plus Coulombic interactions. The full simulation volume contains 6x5x4=120 unit cells, for a total of 480 NM molecules ($N$=3360 atoms). Again local thermal fluctuations are investigated by subdividing into smaller blocks, which are orthorhombic to match the crystalline symmetry. The block size for all NM results presented here is a single unit cell: 1x1x1 in the *axbxc* axis directions, containing 4 molecules, or $<n>$=28 atoms. As with the L-J lattice, each block has a fixed volume and location, filling the entire volume, so that the number of blocks equals the number of unit cells. NM was chosen for study because it provides a system of moderate complexity with realistic potentials, thereby complementing the basic L-J model to establish the generality of our results, and because there are other simulations and experiments with which to compare.

Production runs for the data presented here were obtained with the full system in the microcanonical ensemble (except for one set of data where the Nosé-Hoover thermostat was added as a test). Thus, the atoms were governed solely by Newton's laws without any additional thermal bath. The equilibrium density of the static crystal was determined by adjusting the density at the lowest $T$ until the lattice had zero pressure ($\rho_0$=1.09 for the L-J model), which of course coincides with the minimum *pe*. To maintain isochoric conditions, some simulations used this same $\rho_0$ at higher $T$. Other simulations were initialized using the *NPT* ensemble to minimize pressure at all $T$, which yielded similar results throughout the crystalline phase, establishing that thermal expansion does not significantly alter the behavior. The usual time-step was 0.25 fs for NM and 0.001 L-J units for the L-J model, with most production runs lasting 10,000 time-steps. For L-J modeling of Ar [77,78], using $\sqrt{\frac{6.634 \times 10^{-26} kg}{\varepsilon}} 0.3405 \times 10^{-9} m$ → 2.156 ps/(L-J unit) yields a usual time-step of 2.156 fs. These time-steps were reduced by a factor of 2-10 at higher $T$, and at lower $T$ to verify that there was negligible integration and averaging error. Indeed the total energy was conserved to within a typical deviation of less than a part per million, with no discernable drift. Similarly there was no discernable drift from zero for the average momentum in each direction. Simulations were initialized for at least $10^5$ time-steps before starting each production run. Some simulations were initialized for up to $10^6$ time-steps, with no change in the net behavior, confirming that the systems were well-equilibrated. For the L-J model most data were taken during production runs of 200 time intervals (50 time-steps per interval), so that the total number of distinct replicas for each average ranged between 5400 for $\ell$=8 to more than one million for $\ell$=1 and 2. For NM, most production runs had 1000 time intervals (10 time-steps per interval) yielding a total of 120,000 distinct replicas. For verification, simulations have been repeated independently at different laboratories under a wide range of conditions, including different densities, time-step sizes, and time constants for the thermal bath during initialization, always yielding consistent results.

**IV. Results**

Simulations of the L-J model as a function of the interaction cutoff radius, Fig. 2, establish that 2nd-neighbor interactions are the primary source of excess *pe* fluctuations. Recall that 2nd-neighbor interactions are significant when $r_c>1.6$, while for $1.3 \leq r_c \leq 1.6$ the only interaction is between 1st-neighbor atoms, yielding a robustly harmonic lattice. The solid lines in Fig. 2 show the specific heat from *ke* (cyan) and *pe* (purple), which should equal 3/2 via the equipartition theorem (left side of Eq. (3)). Indeed, from simulations over six temperatures ($kT/\varepsilon$=0.0005 to 0.02) and thirteen cutoff radii ($1.0 \leq r_c \leq 2.5$) the solid cyan line yields d<*ke*>/d($kT$)=1.4992±0.0007, while similar averaging of the solid purple line (whenever the 1st-neighbor interaction is robustly harmonic, $r_c \geq 1.3$) yields d<*pe*>/d($kT$)=1.494±0.002, confirming that these MD simulations give excellent agreement with the expected equipartition of the average energies. (Values slightly below 3/2 can be attributed to anharmonicity in the *pe*, which is more relevant at higher *T*, and from using finite differences instead of derivatives.) For the dynamics, however, whenever $r_c \geq 1.3$ the normalized energy fluctuations clearly differ from the analogous EFR (right side of Eq. (3)). First focus on the *ke* fluctuations (red dotted line), with error bars from averaging over six temperatures ($kT/\varepsilon$=0.0005 to 0.02). These *ke* fluctuations show subtle but significant deviations from the EFR: <$n$><($\Delta ke$)$^2$>/($kT$)$^2$ = 1.356± 0.008 when averaged over $1.7 \leq r_c \leq 2.5$. Much more conspicuous are the anomalies in *pe* fluctuations (open symbols connected by dashed lines), exhibiting two distinct regimes of deviations from the EFR. If the only interaction is effectively harmonic between 1st-neighbor atoms, $1.3 \leq r_c \leq 1.6$, normalized *pe* fluctuations are independent of *T* (as expected from the EFR), but the magnitude of *pe* fluctuations is about 40% below the specific heat: <$n$><($\Delta pe$)$^2$>/($kT$)$^2$=0.93±0.02 when averaged over six temperatures ($kT/\varepsilon$=0.0005 to 0.02). Whereas, if interactions include 2nd-neighbor atoms $r_c>1.6$, the normalized *pe* fluctuations are strongly temperature dependent, diverging as $T \rightarrow 0$. Indeed, energy fluctuations match the specific heat only in the limit of purely repulsive, ideal-gas-like interactions, $r_c<1.3$. The crucial result established by Fig. 2 is that at low *T*, *pe* fluctuations in small blocks are below the expected EFR regardless of the strength of harmonic interaction between 1st-neighbor atoms, with excess *pe* fluctuations appearing abruptly when 2nd-neighbor interactions are added. Thus, these excess *pe* fluctuations cannot come from any simple interface effect, nor from any artifact in simulating the systems, nor from any procedure in analyzing the results.

Solid symbols and right-hand scale in Fig. 2 present various correlations as a function of $r_c$. We define the normalized *pe* correlations between fluctuations in a central block and its $\eta$ equivalent $j^{th}$-neighbor blocks by $C_{pe,j}(t) = \eta \frac{<\Delta pe_0(0)\Delta pe_j(t)>}{<\Delta pe_0(0)\Delta pe_0(0)>}$. (A similar expression gives *ke* correlations, $C_{ke,j}(t)$.) Here $\eta$=1 when *j*=0 (autocorrelations), while when $j \geq 1$: $\eta$=6 for cubic symmetry in the L-J model, with $\eta$=2 for orthorhombic NM. These values of $\eta$ ensure that $C_{pe,j}(t)$ includes the total contribution to energy from all equivalent blocks. Specifically, for the L-J lattice, if all excess energy in a *pe* fluctuation comes solely from 1st-neighbor blocks, then on average 1/6 of this excess *pe* comes from each 1st-neighbor block, yielding $C_{pe,1}(0) = 6(-1/6) = -1$ with no *pe* from other blocks, or from *ke*. Solid squares in Fig. 2 show that the initial (*t*=0) *pe* correlations between 1st-neighbor blocks (*j*=1) are positive ($C_{pe,1}(0)>0$) when the only interaction is robustly harmonic between 1st-neighbor atoms $1.4 \leq r_c \leq 1.6$, coinciding with reduced *pe* fluctuations; while $C_{pe,1}(0)<0$ when interactions include 2nd-neighbor atoms causing excess *pe* fluctuations. Solid circles show the normalized cross-correlation between *pe* and *ke* within each block, which is significantly negative in the robustly harmonic lattice, $C_{kepe,0}(0) = \frac{<\Delta ke_0(0)\Delta pe_0(0)>}{\sqrt{<(\Delta ke)^2><(\Delta pe)^2>}} = -0.047 \pm 0.005$ at $r_c = 1.5$. However, when interactions include 2nd-neighbor atoms ($r_c>1.6$), $C_{kepe,0}(0) = 0.011 \pm 0.009$ is negligible, as needed for canonical-ensemble behavior. Solid triangles show the normalized cross-correlation between *pe* and *n* within each block, which is small in the robustly harmonic lattice ($r_c \leq 1.6$), $C_{pen,0}(0) = \frac{<\Delta pe_0(0)\Delta n_0(0)>}{\sqrt{<(\Delta pe)^2><(\Delta n)^2>}} = 0.014 \pm 0.003$, but jumps up sharply when $r_c>1.6$ and reaches 0.506±0.011 for $r_c>3.0$ (not shown). Thus, excess *pe* fluctuations are correlated with *n* fluctuations and occur only if interactions extend beyond 1st-neighbor atoms, although such long-range interactions seem to be necessary to achieve the canonical-ensemble requirement of independently fluctuating *ke* and *pe*.

Figure 3 (A) presents the temperature dependence of the ratio of distinct contributions to energy $\frac{d<pe>/d(kT)}{d<ke>/d(kT)}$ (solid lines) and their fluctuations $\frac{<(\Delta pe)^2>}{<(\Delta ke)^2>}$ (symbols). We use these ratios to eliminate the possibility of artifacts from finite differences in temperature for the derivatives, and finite differences in time for the fluctuations. Red squares and upper scale show NM, other symbols and lower scale show the L-J model. Thus, the L-J data points in Fig. 3 (A) come from the ratio of *pe* to *ke* behavior shown in Fig. 2, but as a function of *T* at fixed values of $r_c$. For classical particles in the canonical ensemble at low *T*, *pe* and *ke* should give equal contributions to the total energy (left side of Eq. (3)), and its fluctuations (right side of Eq. (3)) [13]. First focus on the solid lines, showing $\frac{d<pe>/d(kT)}{d<ke>/d(kT)}$ from the LJ model for three different values of $r_c$, given by the thickness and color of the line. At low *T*, where each atom oscillates around an effectively harmonic potential in the fcc lattice, the equipartition theorem predicts equal contributions to the specific heat. Indeed, by averaging over five temperatures ($kT/\varepsilon < 0.2$) we find $\frac{d<pe>/d(kT)}{d<ke>/d(kT)} = 0.997 \pm 0.004$, $0.999 \pm 0.004$, and $0.9998 \pm 0.0013$ for $r_c$=1.5, 2.0, and 6.0, respectively. At higher *T*, each solid line has a peak that identifies the melting temperature, which ranges from $kT/\varepsilon$=0.75 to 1.5 for $r_c$=1.5 to 6.0, similar to the values of $kT/\varepsilon$=0.78 to 1.70 found from other simulations at densities of $\rho_0$=0.97 to 1.10 [80]. Above the melting temperature each line drops sharply, as expected in the fluid phase where atoms no longer oscillate about a *pe* minimum. At these high *T*, the symbols show that the ratio of energy fluctuations also drops sharply, $\frac{<(\Delta pe)^2>}{<(\Delta ke)^2>} < 1.0$, consistent with the EFR. However at $kT/\varepsilon << 0.1$, symbols show that $\frac{<(\Delta pe)^2>}{<(\Delta ke)^2>} \gg 1.0$ from microcanonical simulations of small blocks $<n>=32$ (circles) and large blocks $<n>=2048$ (solid diamonds) in much larger volumes containing a total number of atoms $N=55{,}296$ (open), or $N=442{,}368$ (solid), from fluctuations of individual atoms (cyan stars), and from canonical simulations using the Nosé-Hoover thermostat (open diamonds). Similarly, red squares show that NM has $\frac{<(\Delta pe)^2>}{<(\Delta ke)^2>} = 5.9$ at 500 K, and 3000 at 0.1 K.

The broken lines in Fig. 3 (A) show that excess *pe* fluctuations diverge inversely proportional to temperature at low *T*. Specifically, $\frac{<(\Delta pe)^2>}{<(\Delta ke)^2>} = (232\ \mathrm{K})/T$ for NM (red dashed line), while for the L-J model the divergent part of the *pe* fluctuations can be approximated by the empirical expression $(0.038\pm0.003)\,\varepsilon/(kT\ell^{(0.75\pm0.05)})$ (black dotted lines). Assuming that this size dependence extends to much larger simulations, reducing the excess fluctuations to within 10% of the EFR at $kT/\varepsilon=0.001$ requires blocks of size $\ell=2750$, corresponding to more than 83 billion atoms. Although uncertainty in the exponent of $\ell$ allows for the possibility that excess fluctuations could vary as $1/\ell^{0.67}$, the behavior cannot be a simple surface effect in the dominant 1st-neighbor interaction because solid triangles show that the excess *pe* fluctuations arise abruptly when the cutoff radius is increased to include 2nd-neighbor interactions. Specifically, at low *T* as $r_c$ increases from 1.5 (down triangles) to 2.0 (up triangles) the ratio $\frac{<(\Delta pe)^2>}{<(\Delta ke)^2>}$ increases by an order of magnitude due to the onset of the $1/T$ divergence, while the ratio $\frac{d<pe>/d(kT)}{d<ke>/d(kT)}$ is unchanged (solid lines). Again, although MD simulations accurately yield expected average energies from the left side of Eq. (3), fluctuations show significant deviations from the right side of Eq. (3).

Figure 3 (B) shows the initial *pe* correlations between neighboring blocks. For NM (red squares), this $C_{pe,1}(0)$ reaches $-0.96$ at 0.1 K. Thus, at 0.1 K, 96% of the *pe* in a fluctuation of each block ($<n>=28$ atoms) comes from the *pe* of its two neighboring blocks along the *a*-axis. In the L-J model, when $r_c=6.0$ (black circles, with error bars from averaging over blocks containing $<n>=106\text{-}2048$ atoms) $C_{pe,1}(0)$ varies from 0.067 at high *T* (black line) to $-0.72$ at $kT/\varepsilon = 0.0005$. Comparison to Fig. 3 (A) reveals that the onset of excess *pe* fluctuations coincides with the onset of negative $C_{pe,1}(0)$, and that the amplitude of excess *pe* fluctuations increases monotonically as $C_{pe,1}(0)$ becomes increasingly negative. Similarly for $r_c=2.0$ (up triangles), the onset of excess *pe* fluctuations in (A) coincides with the onset of negative $C_{pe,1}(0)$ in (B), while for $r_c=1.5$ (down triangles) this $C_{pe,1}(0)=0.21\pm0.03$ (solid line) shows no tendency towards negative values down to the lowest *T*. Negative values of $C_{pe,1}(0)$ indicate that an energy increase in the central block is at least partially compensated by an energy decrease in neighboring blocks, altering the net energy that reaches distant parts of the simulation volume that serve as the large heat reservoir.

Figures 4 (A-C) show the time dependence of *pe* correlations between blocks containing a single unit cell, from simulations of the L-J model (A & B) at $kT/\varepsilon$=0.0005 and NM (C) at *T*=100 K. Line color gives the distance between blocks: *j*=0 (black) and *j*=1 (red), with *j*=2 (green) and *j*=3 (blue) for more-distant blocks in (C). All autocorrelations (*j*=0) approach 1.0 as $t \rightarrow 0$ due to the normalization. For *t*>0, black and red lines exhibit damped harmonic oscillations that remain roughly $180^0$ out-of-phase in (A) and (C), and roughly in-phase in (B), perpetuating the initial sign of the correlation. Indeed, (A) and (C) show that long-range interactions have negative $C_{pe,0}(t)*C_{pe,1}(t)$, with roughly constant $C_{pe,0}(t)+C_{pe,1}(t)$, indicating that excess *pe* is traded back and forth between neighboring blocks, characteristic of short-wavelength (zone-boundary) motion. In contrast, (B) shows that a robustly harmonic lattice has positive $C_{pe,0}(t)*C_{pe,1}(t)$. The green and blue lines in (C) show negligible initial correlations between more-distant blocks, indicating that the energy is initially localized. Eventually these distant correlations increase with increasing *t*, then finally decrease, characteristic of a damped wave as excess *pe* from the fluctuation flows out from the central block. The speed of this wave is obtained from the inset of (C) showing distance to the center of increasingly distant blocks (*j**0.522 nm) as a function of time at which each correlation reaches its first maximum. The slope of this line yields 1.11 nm/ps (1110 m/s), somewhat slower than the speed of sound from other simulations (1633 m/s) [81], and from measurements on liquid NM (1300 m/s) [82], which sets the maximum speed of energy dispersal in such insulating substances.

Solid lines in Fig. 4 (D) show time-dependent *ke* correlations in NM, from autocorrelations (black) and 1st-neighbor blocks (red). In contrast to the corresponding $C_{pe,j}(t)$ in Fig. 4 (C), $C_{ke,j}(t)$ relaxes monotonically, and much slower. Broken lines show fits to the *ke* autocorrelation: exponential at short times $C_{ke,0}(t) \propto e^{-t/\tau}$ (dotted purple) with a time constant of $\tau$=0.41±0.02 ps, and stretched-exponential at longer times $C_{ke,0}(t) \propto e^{-(t/\tau)^\beta}$ (dashed magenta) yielding $\tau$=16.5±0.2 ps and $\beta$=0.73±0.01. Previous simulations of NM after excess *ke* was added to a central molecule find similar timescales, $\tau$=11.6-13.6 ps, but with $\beta$=1 [83]. The two types of response we find for $C_{ke,0}(t)$ suggest that the short-time behavior comes from localized dynamics, similar to $C_{pe,j}(t)$, while slower relaxation involves transfer of *ke* to the large heat reservoir, similar to the behavior found in simulations of copper [84]. Measurements of temperature as a function of time after shock-induced ignition in PETN (triangles), and other energetic materials analogous to NM [85], also show a crossover from exponential relaxation at short times to non-exponential relaxation at longer times, somewhat similar to our simulations of $C_{ke,0}(t)$. However, to match the simulations, the measurements were offset by a final temperature and normalized by an initial temperature, with the time scale reduced by a factor of 100,000. This large difference in time scales could be due to the difference in length scales, or another mechanism, emphasizing the need for further study.

Figure 5 compares frequency-dependent correlations in the energies, and density, of the L-J model as the interaction cutoff radius is changed from including 2$^{nd}$-neighbors $r_c$ = 2.0 (A) to only 1$^{st}$-neighbors $r_c$ = 1.5 (B). The frequency scale (in THz) is set by the parameters for Ar (2.156 ps/L-J unit). Symbols come from the Fourier transform of normalized autocorrelations in atom density (green triangles), *ke* (red circles), and *pe* (black squares), with blue squares from correlations in *pe* between 1$^{st}$-neighbor blocks. Fluctuations in *n* are obtained from simulations having block walls aligned with the equilibrium atomic planes, while other simulations have block walls shifted to maintain the exact number of atoms *n*=108 (solid) or *n*=864 (open). Data come from averaging spectra at three temperatures ($kT/\varepsilon$ = 0.0005 to 0.002, justified at low-*T* for effectively harmonic modes), with error bars for the solid squares that are visible only if larger than the symbol size. All data are obtained from various correlations as a function of time. For example, the black squares in Figs. 5 (A) and (B) come from the Fourier transform of the time-dependent *pe* autocorrelations, similar to those shown by the black lines in Figs. 4 (A) and (B), respectively. Specifically, the *pe* correlations come from the real part of the discrete Fourier transform, $\sum_{t=-\theta}^{\theta}[C_{pe,j}(|t|)]\cos\left(\frac{2\pi f t}{4\theta}\right)$, with the quantity in square brackets replaced by $C_{ke,0}(|t|)$ or $C_{n,0}(|t|)$ for correlations in *ke* or *n*, respectively. The Wiener-Khinchin theorem implies that the autocorrelations (*j*=0) yield power-spectral densities, so that separate peaks identify distinct normal modes. Arrows in (A) indicate how various modes in the fluctuations of *n* and *pe* shift when the linear size of each block is doubled, $\ell$=3→6. Key results are found by contrasting the behavior between (A) and (B). First note that only in (A) do *pe* correlations between 1$^{st}$-neighbor blocks (blue) show modes that are roughly inverse of the autocorrelations (black), consistent with negative correlations shown in Figs. 2, 3 (B), and 4 that appear abruptly when interactions extend beyond 1$^{st}$-neighbor atoms. Second, in Fig. 5 (B) the *pe* spectra (black squares) and *ke* spectra (red circles) show strong similarities, whereas *pe* and *n* (green triangles) are more similar in (A). Finally note that only in (A) do the characteristic frequencies and total number of conspicuous *pe* modes depend on the size of the block, as indicated by the black arrows. This size dependence from (A) is depicted in the inset of (B), where solid symbols show how various *pe* modes depend on 1/$\ell$. Open symbols in this inset show similar behavior for atom-density modes, except for the low-frequency mode that decreases as 1/$\ell$→0 (see also the green arrow in (A)), opposite to all *pe* modes.

Figure 6 shows energy distributions of the L-J model at $kT/\varepsilon$ = 0.001. The distributions are from histograms for blocks containing a single unit cell, *n*=4.00. Symbols show that *pe* (open) and *ke* (solid) have similar distributions if $r_c$=1.5 (squares), but starkly different distributions if $r_c$=2.0 (circles). (Note that *pe* has been offset by *pe*$_g$, the ground-state energy as *T*→0 as deduced by the equipartition theorem, <*pe*>–*pe*$_g$=<*ke*>.) Lines show fits to the data using the canonical-ensemble density of states [31], $(AE/kT)^\gamma e^{-BE/kT}$ (which gives significantly better agreement than a Gaussian distribution, not shown), indicative of thermal-equilibrium behavior of finite-sized systems. Here *A* and *B* are dimensionless factors that scale the energy

*E*. For an ideal gas the exponent is $\gamma=3n/2-1=5.00$. The dashed line shows a fit to the *ke* distributions yielding $\gamma=4.99\pm0.06$. The solid lines yield $\gamma=7.5\pm0.1$ for $r_c=1.5$ (cyan) and $\gamma=325\pm1$ for $r_c=2.0$ (blue). Here, $\gamma>5$ indicates that the effective degrees of freedom significantly exceed those expected for an ideal gas of *n* atoms, attributable to interactions with atoms outside the block. In other words, interactions cause the effective degrees of freedom for *pe* fluctuations in each block to exceed those of its atoms, especially for long-range interactions at low *T* where correlations between *pe* in neighboring blocks is strong and negative.

**V. Discussion**

The results presented here confirm that classical MD simulations of diverse models exhibit energy fluctuations that clearly deviate from standard statistical mechanics. Indeed, Figs. 2 and 3 show that *pe* fluctuations in nanometer-sized blocks at low *T* can deviate from the expected EFR by an order of magnitude or more. We now discuss various mechanisms and interpretations that could cause these excess fluctuations.

**A) Anomalous *pe* fluctuations come from localized energy**

Figures 2-5 establish that excess *pe* fluctuations inside a central block occur only when there are negative correlations with the *pe* in the surrounding shell of similar blocks. From Fig. 4 (C) it can be seen that a vanishingly small fraction of the excess *pe* in a local fluctuation reaches distant blocks before the excess *pe* returns to zero. Thus, the thermal bath governing *pe* fluctuations in each system comes primarily from its 1st-neighbor blocks, yielding an explicit local bath that is comparable in size to the system itself; not a large heat reservoir.

The clear connection between energy localization and deviations from the EFR can be characterized by the net energy that reaches the heat reservoir. Specifically, let the *PE* of the system (central block of *n* atoms) fluctuate by a total amount $\Delta PE=n\Delta pe$, so that the concurrent change in *PE* of blocks at distance $j\ell$ is $C_{pe,j}(0)\Delta PE$. For simplicity consider only the dominant correlation from 1st-neighbor blocks, so that the net change for *PE* fluctuations can be approximated by $(1+C_{pe,1}(0))\Delta PE$, yielding an effective Boltzmann's factor of $e^{-(1+C_{pe,1}(0))\Delta PE/kT}$. Equivalently, the fluctuations can be treated as if each block has an effective temperature for *pe* fluctuations of: $T'=T/(1+C_{pe,1}(0))$, where $C_{pe,1}(0)<0$ yields $T'/T>1$. For example, from the black circles and red squares in Fig. 3 (B) at the lowest *T*, $C_{pe,1}(0) = -0.72$ and $-0.96$ gives $T'/T=3.6$ and 25 for the L-J model and NM, respectively. In contrast, when the only interaction is robustly harmonic between 1st-neighbor atoms, $C_{pe,1}(0) = +0.21$ (solid line at low *T* in Fig. 3 (B)) yields a reduced effective temperature, $T'/T=0.83$. Fluctuations from an effective local temperature were used by Einstein in 1910 to describe critical opalescence [86]. More recently, an effective local temperature analogous to our $T'$ has been used to describe ring statistics in amorphous layers of $SiO_2$ [87], also with $T'/T<1.0$ implying positive correlations, $C_{pe,1}(0)>0$. Another effective temperature can be deduced from the *pe* fluctuations using the right-side of Eq. (3): $T''/T=\sqrt{<(\Delta pe)^2>/<(\Delta ke)^2>}$. For $r_c=1.5$, *pe* fluctuations in the effectively harmonic lattice, $<(\Delta pe)^2>/<(\Delta ke)^2>=0.651$, yield $T''/T=0.807$, within 3% of the value from *pe* correlations. Similarly, for $r_c=2.0$ (up triangles) at the lowest $T$, $C_{pe,1}(0)=-0.5828$ yields $T'/T=2.397$ while $<(\Delta pe)^2>/<(\Delta ke)^2>=5.700$ yields $T''/T=2.387$. Thus, adding the *pe* from 1st-neighbor blocks accurately defines the effective local temperature for interactions that are relatively short-range: $T'<T$ for 1st-neighbor interactions, and $T'>T$ when 2nd-neighbor interactions are added. Furthermore, for $r_c=6.0$ (black circles and red squares) at the lowest $T$, $<(\Delta pe)^2>/<(\Delta ke)^2>=52$ and 3000 gives $T''/T=7.2$ and 55 for the L-J model and NM, respectively. Although these values of $T''$ are about twice the values of $T'$, full quantitative agreement is not expected due to additional contributions to $T'$ from more-distant blocks, and from ambiguity in deciding where the local bath ends and the large heat reservoir begins. Nevertheless, the magnitude and sign of *pe* correlations shown in Fig. 3 (B) control the deviations from the EFR shown in Fig. 3 (A), so that the expected increase (or reduction) in *pe* fluctuations can be estimated by knowing the net energy that reaches the large heat reservoir needed for Boltzmann's factor.

Because most models of interacting particles will have correlations that influence the net energy during fluctuations, modifications of Boltzmann's factor may often be necessary to describe the local dynamics. The correction can be approximated by the total energy that reaches the heat reservoir due to direct interactions between the system and surrounding shell of similar systems $\Delta PE \rightarrow (1+C_{pe,1}(0))\Delta PE$, or by an effective local temperature $T^*=T/(1+C_{pe,1}(0))$. Both interpretations retain the usual Boltzmann's factor for the statistics of fluctuations by simply combining the change in energy of the central block with the change of its neighbors. A drawback is that $(1+C_{pe,1}(0))\Delta PE$ comes not only from the system, but also from its environment, which must be measured separately or otherwise approximated. An alternate approach that could allow refocusing solely on the system is to include a nonlinear correction from the local entropy. Indeed, a quadratic correction to Boltzmann's factor in the Metropolis algorithm for MC simulations alters the *pe* fluctuations so that they are consistent with entropy that is extensive and additive [28]. Furthermore, this correction significantly improves agreement between MC simulations of standard models and the measured response of many materials, including critical fluids and ferromagnets [29-33]. Future work will be necessary to determine whether the failure of the EFR for MD simulations may also be treated by a nonlinear correction from the local entropy, without having to include explicit information from surrounding parts of the sample.

### B) Excess *pe* fluctuations come primarily from 2nd-neighbor interactions

Figures 2, 3, 5, and 6 establish that excess *pe* fluctuations in the L-J model require interactions that extend beyond 1st-neighbor atoms. Thus, a likely mechanism comes from an incipient instability due to the attractive force between 2nd-neighbor atoms. This attractive force tends to favor a type of distortion that can bring 2nd-neighbor atoms closer together, without changing the distance between 1st-neighbors. Specifically, consider three atoms that are initially at the vertices of an isosceles right triangle, somewhat like the symbol Λ, as in the face of an fcc unit cell. Now imagine moving the 2nd-neighbor atoms towards each other to form an equilateral triangle, Δ. If the equilibrium distance between 1st-neighbor atoms is preserved, the net energy of the 2nd-neighbor atoms can be reduced by about 77% (from $-\frac{15}{64}\varepsilon$ to $-\varepsilon$). Stability against such distortions comes from other atoms in the lattice that maintain the net fcc structure, consistent with the negative correlations in the *pe* of neighboring blocks as shown in Figs. 2-5. Positive correlations between fluctuations in *pe* and *n* shown in Figs. 2 and 5 (A) imply that the Δ-like distortion causes additional atoms to move out of the block through the surfaces perpendicular to the axis of the distortion. Direct evidence for the importance of this Δ-like distortion comes from L-J simulations, and measurements of Ar clusters, confirming that an icosahedral structure is preferred for systems of less than ~3000 atoms [88].

Figure 5 reveals several details about the excess *pe* fluctuations, their connection to atomic motion, and why they are absent if the interaction cutoff range is reduced from including 2nd-neighbors $r_c = 2.0$ (A), to only 1st-neighbors $r_c = 1.5$ (B). Fluctuations in *n* (green triangles) show a large-amplitude low-frequency mode, and multiple mid-frequency modes, all of which simply shift to lower frequencies as $r_c$ is reduced. Fluctuations in *ke* (red circles) show no mid-frequency modes, but with high-frequency modes that also simply shift to lower frequencies as $r_c = 2.0 \rightarrow 1.5$, again with no obvious change in the general features of the modes. In contrast, *pe* fluctuations (black squares) are fundamentally altered by the addition of 2nd–neighbor interactions, from closely matching *ke* fluctuations if $r_c = 1.5$ to matching most of the density modes if $r_c = 2.0$. This abrupt switch in correlation for *pe* fluctuations from *ke* to *n* is confirmed on the right side of Fig. 2 by the shift from nonzero $<pe_0 ke_0>$ (solid circles) for $r_c \leq 1.6$, to nonzero $<pe_0 n_0>$ (solid triangles) for $r_c > 1.6$. Thus, atomic vibrations that cause excess fluctuations couple strongly to the *pe* only if interactions extend beyond 1st-neighbor atoms. As expected: the atomic motion is weakly coupled to 1st-neighbor interactions that are robustly harmonic (near the bottom of a potential well), whereas this motion couples strongly to 2nd-neighbor interactions that are intrinsically anharmonic (on the side of the potential well).

**C) Excess *pe* fluctuations involve zone-boundary vibrational modes that are localized**

Symbols in the inset of Fig. 5 (B) show that all peaks from (A) for both *pe* (solid) and *n* (open) vary with the inverse linear dimensions of the block. We use the known dispersion relations of the L-J model (and its similarity to argon) [78,89,90] to assign a specific vibrational mode to each peak. The low-frequency mode (found only in the *n* fluctuations) that initially increases linearly with increasing $1/\ell$ and reaches a maximum at $1/\ell = 0.5$ can be characterized by $f(\text{THz}) = (0.557 \pm 0.008) \sin(\pi/\ell)$ (dashed line). The behavior of this mode is consistent with long-wavelength transverse vibrations in the (1,1,1) direction from the Γ- to L-points. *pe* fluctuations do not couple strongly to this mode because 2nd-neighbor atoms tend to move in-phase for long-wavelength motion, balancing the anharmonic terms. All other modes extrapolate to non-zero frequencies as $1/\ell \rightarrow 0$, characteristic of short-wavelength zone-boundary modes that can be characterized by a quadratic dependence, $f$ (THz)$=A_i(1/\ell)^2+C_i$ (solid lines) where $A_i$ is a constant that governs the size dependence of the $i^{th}$ mode, and $C_i$ is its characteristic frequency. In units of THz we find $C_0$=1.63±0.02, $C_1$=1.34±0.04, $C_2$=1.10±0.03, $C_3$=0.84±0.02, and $C_4$=0.59±0.04. The ratio $C_0/C_2$=1.483 accurately matches the ratio for longitudinal- to transverse-mode frequencies of 1.480 at the X-point [91]. Although our absolute $C_0$ is about 12 % below their value of 1.85 THz, the difference can be attributed to different cutoff radii, and their choice of energy and length scales for Ar atoms ($\varepsilon/k$=135 K and 0.398 nm), which alters the characteristic frequencies. The $C_1$ frequency is most consistent with the longitudinal mode near the K-point. The $C_3$ and $C_4$ frequencies have no clear connection to zone boundary modes. Therefore, we attribute these to transverse vibrations with longer wavelengths that nest within larger blocks, consistent with the fact that they do not occur in smaller blocks. Note that the characteristic frequencies of all three transverse modes ($C_2$-$C_4$) decrease relatively rapidly with increasing $1/\ell$, perhaps because they are governed by shear motion that is softer and more-strongly dependent on the fraction of surface atoms.

Thermal properties at low $T$ are often calculated by assuming effectively-harmonic plane-wave normal modes. Our analysis involves real-space blocks that show how decorrelations occur with increasing distance between blocks. Correlations are usually expected to diminish monotonically with distance, as we find for robustly harmonic interactions. However, when $r_c$>1.6 we find strong anti-correlations between neighboring blocks indicative of wave-like modes that are spatially localized. Anderson localization is an established mechanism for localization of electron wavefunctions [48], which has been extended to treat other systems [49], including vibrational modes governed by Hessian matrices [92-94]. Key parameters are the width in the energy distribution ($W$), and the critical width ($W_C$). Normal modes are localized when $W$>$W_C$, and delocalized when $W$<$W_C$. At $kT/\varepsilon$ = 0.001, Fig. 6 shows that $W$ (full width at half maximum) increases by a factor of five when the cutoff radius is increased from $r_c$=1.5 to 2.0. Evidently, 2$^{nd}$-neighbor interactions broaden $W$ sufficiently to localize the normal modes at low $T$. We argue that such broad *pe* distributions arise when Δ-like distortions induce icosahedral-like local structures, with most energies outside the dispersion band found for delocalized plane waves in the bulk fcc lattice, consistent with Fig. 6. Thus, even ideal crystals may have $W$>$W_C$ for vibrational modes, attributable to randomness in the instantaneous positions of 2$^{nd}$-neighbor atoms. Future work will be necessary to clarify the relevance of Anderson localization for MD simulations of simple systems, and to determine specific values of $W_C$.

### D) Theory for excess *pe* fluctuations

A general theory for excess *pe* fluctuations comes from treating 1$^{st}$- and 2$^{nd}$-neighbor interactions using different thermodynamic ensembles. All ensembles yield equivalent results for static properties of large systems, but the correct ensemble is required for fluctuations of finite-sized systems [16,95]. For example, fast degrees of freedom that do not have time to couple to the large heat reservoir are isolated and must conserve energy locally, while slow degrees of freedom are isothermal. In general there are two types of isolated ensembles: microcanonical with fixed energy if the system is fully isolated, and adiabatic with all energies equally likely for blocks that conserve energy locally via strong coupling to neighboring blocks, effectively isolating the central block from the large heat reservoir. Although our theory for MD simulations is superficially similar to the multicanonical ensemble [96,97] and related to the hot solvent-cold solute problem [98-100], here we use different ensembles concurrently for distinct degrees of freedom. The adiabatic ensemble is appropriate for 2$^{nd}$-neighbor interactions that connect strongly to high-frequency zone-boundary modes, with the canonical ensemble for 1$^{st}$-neighbor interactions that couple to low-frequency zone-center modes. For classical MD simulations, the equipartition theorem ensures that all modes have equal average energy at all $T$. In real crystals at low $T$, quantum effects reduce the influence of high-frequency modes, but significant anharmonic effects remain even when $T \rightarrow 0$ as evidenced by finite thermal conductivity and residual thermal expansion [101].

At sufficiently low $T$ most 1$^{st}$-neighbor potentials are robustly harmonic, $pe_1(\delta) \approx \frac{1}{2}a\delta^2$; while the pair-potential for 2$^{nd}$-neighbor atoms is dominated by a linear (anharmonic) term, $pe_2(\delta) \approx -b\delta$. Here $\delta$ is a relative (dimensionless) displacement, so that constants $a$ and $b$ have units of energy. To be specific, again consider three atoms in the L-J model (fcc lattice) that are initially at the vertices of the isosceles triangle, aligned so that the lower two atoms are on the $x$-axis, with $r_2=\sqrt{2}r_0$ the equilibrium (2$^{nd}$-neighbor) distance between these two atoms, and $r_0=2^{1/6}$ the equilibrium (1$^{st}$-neighbor) distance to the upper (face-centered) atom. Assume that only the lower-left atom moves, constrained to stay on the $x$-axis, with the origin ($x=0$) at the center of its motion. Let the distance between the 2$^{nd}$-neighbor atoms be reduced to $r_2(1-\delta)$ due to a fluctuation, so that to lowest order the change in their pair-potential is $pe_2(\delta) = \frac{15}{128}\varepsilon + 2\varepsilon \left[ \left(\frac{1}{r_2(1-\delta)}\right)^{12} - \left(\frac{1}{r_2(1-\delta)}\right)^{6} \right] \approx -\varepsilon \frac{21}{32}\delta$. (Note that the potential is divided by 2 to yield the $pe$ of each atom.) The 1$^{st}$-neighbor atom is at an angle of 45$^0$ above the $x$-axis, so that for small displacements the 1$^{st}$-neighbor distance becomes $r_0(1-\cos(45^0)r_2\delta/r_0) = r_0(1-\delta)$. The lowest-order change in the 1$^{st}$-neighbor potential is $pe_1(\delta) = \frac{1}{2}\varepsilon + 2\varepsilon \left[ \left(\frac{1}{r_0(1-\delta)}\right)^{12} - \left(\frac{1}{r_0(1-\delta)}\right)^{6} \right] \approx \frac{1}{2}36\varepsilon\delta^2$. Thus, the total change in $pe$ for the single atom at the origin due to its displacement is: $pe_{1+2}(\delta) = (\frac{1}{2}a\delta^2 - b\delta)$ with $a=36\varepsilon$ and $b=\frac{21}{32}\varepsilon$.

To maintain generality we start with a mixture of adiabatic and canonical ensembles for the 2$^{nd}$-neighbor interactions. Specifically, let the energy reaching the large heat reservoir from the linear term be $-c\delta$ (with $c$ constant), so that Boltzmann's factor becomes $g(\delta) = e^{-(\frac{1}{2}a\delta^2 - c\delta)/kT}$. The average energy is:

$<pe_{1+2}> = \frac{\int_{-\infty}^{\infty} pe_{1+2}(\delta)g(\delta)d\delta}{\int_{-\infty}^{\infty} g(\delta)d\delta} = \frac{1}{2}kT + \frac{c}{2a}(c-2b)$, yielding the equipartition theorem for a single atom ($n=1$) moving in 1-dimension, and a constant energy reduction for $b>c/2$. Thus, integrating directly over $d\delta$ with $g(\delta)$ (instead of over $dE$ with $p_E g_E$) provides an alternative way to obtain the left side of Eq. (3), and the heat capacity is unchanged by linear terms in $pe(\delta)$. Although $<pe_{1+2}>$ is minimized by the canonical value $c=b$, this involves a static displacement from the origin that is inhibited by other neighbors in the fcc lattice. The energy fluctuations become: $<(\Delta pe_{1+2})^2> = \frac{\int_{-\infty}^{\infty}[pe_{1+2}(\delta)]^2 g(\delta)d\delta}{\int_{-\infty}^{\infty} g(\delta)d\delta} - <pe_{1+2}>^2 = \frac{1}{2}(kT)^2 + \frac{(b-c)^2}{a}kT$. Hence, $\frac{<(\Delta pe_{1+2})^2>}{(kT)^2} = \frac{1}{2} + \frac{(b-c)^2}{akT}$ yields a term that agrees with the right side of Eq. (3), plus a term that diverges as $1/T$, qualitatively consistent with the dashed lines in Fig. 3 (A); but only if $c \neq b$. Quantitatively, using $c=0$ for a purely adiabatic ensemble, with $\frac{<(\Delta ke)^2>}{(kT)^2} = \frac{1}{2}$ for motion in one dimension, yields $\frac{<(\Delta pe_{1+2})^2>}{<(\Delta ke)^2>} - 1 = \frac{2b^2}{akT} = 0.0239\frac{\varepsilon}{kT}$, somewhat less than the value of $0.038\frac{\varepsilon}{kT}$ for $\ell=1$ found in Fig. 3 (A). A better model for the fcc lattice has 12 1$^{st}$-neighbor atoms yielding $a \rightarrow 12a$, and 6 2$^{nd}$-neighbors. If these 2$^{nd}$-neighbor atoms are precisely equidistant from the central atom then $b \rightarrow 0$, valid for static equilibrium and long-wavelength motion. For local fluctuations, however, a more accurate value comes from the root-mean-squared displacements of the 2$^{nd}$-neighbor atoms, yielding $b \rightarrow 6\frac{b}{\sqrt{2}}$ and $\frac{(6b)^2}{12akT} = 0.0359\frac{\varepsilon}{kT}$, well within the uncertainty for the $\ell=1$ behavior deduced from Fig. 3 (A). Although this basic model yields behavior consistent with the simulations, a detailed theory must include normal modes for all atoms and interactions, which is a goal for future studies.

Despite its simplicity, the three-atom model exhibits several features found in the simulations. The model shows accurate equipartition of the average $pe$, whereas $pe$ fluctuations diverge proportional to $1/T$, consistent with the dashed lines in Fig. 3 (A). The model often lowers the average $pe$ of the three atoms, at the expense of increased $pe$ from surrounding atoms, consistent with negative correlation of $pe$ in neighboring blocks shown in Figs. 2-5. Reduction of $pe$ in the model involves changes in local density, consistent with the sharp onset of correlations in $<pe_0 n_0>$ when $<pe_0 pe_1>$ becomes negative in Fig. 2, and the similarity in high-frequency modes from $pe$ and $n$ fluctuations shown by the black and green symbols in Fig. 5 (A). Also from Fig. 5 (A) the excess $pe$ fluctuations are coupled to high-frequency zone-boundary modes, consistent with the high-frequency oscillations in $pe$ shown in Figs. 4 (A) and (C). Excess $pe$ fluctuations in the model require anharmonic interactions from the 2$^{nd}$-neighbor atom, consistent with the sharp onset of excess $pe$ fluctuations and negative correlations in $<pe_0 pe_1>$ that coincide with the onset of 2$^{nd}$-neighbor interactions shown in Fig. 2. Thus, $\Delta$-like distortions that transiently reduce the local $pe$ are the likely cause of excess $pe$ fluctuations in the simulations. Indeed, using Ovito visualization software [102] to analyze atom positions of the L-J model reveals that there is a small but significant increase in the relative variance of 2$^{nd}$-neighbor distances with the onset of 2$^{nd}$-neighbor interactions, 1.023→1.044±0.003. No such increase in variance is found for the distances to other neighbors, confirming that 2$^{nd}$-neighbor interactions dominate the excess *pe* fluctuations in the simulations. Finally, quantitative agreement with excess *pe* fluctuations requires that no anharmonic term occurs in Boltzmann's factor, $c$=0. The resulting adiabatic ensemble is consistent with an altered local temperature for the excess *pe* fluctuations, as well as the strong negative correlations between 1$^{st}$-neighbor blocks shown in Figs. 2-4, and 5 (A). In any case, the finite speed of energy flow shown in Fig. 4 (C) confirms that local changes in *pe* do not immediately reach the large heat reservoir needed for Boltzmann's factor.

### E) Significance of excess *pe* fluctuations

One way to establish the generality of our results is to review the various models we have found to yield excess *pe* fluctuations, and some of the extreme measures needed to reduce these fluctuations. Figure 3 shows that excess *pe* fluctuations are exhibited by the NM model at most *T*, and by the 12-6 L-J model at low *T*. Although not presented here, excess *pe* fluctuations are also found in the 9-6 L-J model at low *T*, whereas *pe* fluctuations are reduced below the EFR for a lattice of atoms with purely harmonic bonds. Thus, all models we have studied with interactions that extend beyond 1$^{st}$-neighbor atoms exhibit excess *pe* fluctuations, and this excess tends to increase with increasing interaction range. In fact, the key ingredient for excess *pe* fluctuations is that the pair potential must encompass a second type of neighbor that is not at a robustly harmonic minimum. Indeed, Figs. 2 and 3 show that *pe* fluctuations are reduced below the EFR if the cutoff radius of the L-J interaction is reduced to include only the robustly harmonic interactions between 1$^{st}$-neighbor atoms, $1.3 \leq r_c \leq 1.6$. Adding a harmonic tether to each atom also tends to reduce the *pe* fluctuations (not shown), but for the L-J model at low *T*, reducing the excess *pe* fluctuations to within 10% of the EFR requires a harmonic tether that is at least 1000 times stronger than the underlying L-J interaction. Another unrealistic way we have found to reduce *pe* fluctuations is to add a Gaussian potential to each atom that effectively cancels the net force between 2$^{nd}$-neighbor atoms, confirming our observation that excess *pe* fluctuations are due primarily to these 2$^{nd}$-neighbor interactions.

The strongest verification of our results comes from changing the range of interactions. Figures 2 and 3 show that excess *pe* fluctuations in the L-J model decrease with decreasing cutoff radius, dropping abruptly to below the EFR when $1.3 \leq r_c \leq 1.6$, where the only interactions are robustly harmonic between 1$^{st}$-neighbor atoms. Thus, excess *pe* fluctuations cannot come from the 1$^{st}$-neighbor interactions that dominate the magnitude of *pe,* nor from any details in the simulation or data analysis. Specifically, from Fig. 2 at the lowest *T*, a subtle increase in $r_c$ from 1.6 to 1.7 causes stark changes in the *pe* fluctuations: from a constant value that is well below the EFR to a *T*-dependent value that diverges as *T*→0. Meanwhile, the right side of Fig. 2 shows that $<pe_0 ke_0>$ changes abruptly, from significantly negative for $r_c \leq 1.6$ to $<pe_0 ke_0> \approx 0$ for $r_c \geq 1.6$. Clearly, excess *pe* fluctuations are not caused by correlations between *pe* and *ke*. Furthermore, 2nd-neighbor interactions seem to be necessary to yield the canonical-ensemble requirement of $\langle pe_0 ke_0 \rangle \approx 0$, yet the same interactions cause the excess *pe* fluctuations that violate the EFR. Meanwhile, $\langle pe_0 n_0 \rangle$ changes abruptly from near zero for $r_c \leq 1.6$ to significantly positive for $r_c > 1.6$, consistent with Fig. 5 where high-frequency zone-boundary modes couple strongly to the *pe* only if there are 2nd-neighbor interactions. Finally, $\langle pe_0 pe_1 \rangle$ changes abruptly from positive to negative, consistent with the change from reduced to increased effective local *T* in Boltzmann's factor, and again establishing that energy localization measured by these correlations controls how *pe* fluctuations deviate from the standard EFR.

### F) Adding heterogeneity to MD simulations

Figure 6 shows that for $r_c=2.0$ at $kT/\varepsilon = 0.001$ there is a very broad *pe* distribution at low *T*, with many blocks having *pe* below the ground-state energy. Indeed, the fraction of blocks with $pe<pe_g$ is about 31.2% when $r_c=2.0$ and 0.01% when $r_c=1.5$, consistent with an incipient Δ-like distortion that lowers the local energy when the interaction includes 2nd-neighbor atoms. Thus, finite-sized blocks tend to fluctuate towards their preferred icosahedral local structure, as found for finite-sized clusters [88]. Maximum reduction in the local *pe* is inhibited by the increased energy of the surrounding sample, consistent with the negative correlation between neighboring blocks shown in Figs. 2-5. Thus, an appropriate mechanism that transiently decorrelates neighboring blocks would reduce the average *pe* of all blocks. Many experimental techniques have shown that most materials have a mechanism that decorrelates neighboring regions [60-69], unlike the uniform correlations in MD simulations of homogeneous models. These nanometer-sized regions form an ensemble of independent small systems, requiring nanothermodynamics to properly describe each small system that is in thermal contact with an ensemble of similarly small systems [15,16]. Moreover, a heterogeneous distribution of independently relaxing regions provides an intrinsic mechanism for coarse-grained dynamics, which was already recognized by Gibbs as a way to obtain the irreversible 2nd-law of thermodynamics from the reversible microscopic laws of physics [103].

Adding heterogeneity to future MD simulations should facilitate a more-realistic description of thermal and dynamic response. This heterogeneity will: reduce high-energy interfaces between neighboring regions, lower the interaction energy at a given temperature, and restore energy fluctuation relations for the dynamics. To be consistent with experiments, the mechanism must yield distinct regions that fluctuate independent of neighboring regions. For the L-J model, heterogeneity could arise from the long range ($1/r^6$) van der Waals interaction, which requires correlated fluctuations between induced dipoles (London dispersion force). Maintaining this correlation within nanometer-sized regions, while breaking the correlation between neighboring regions, would yield thermodynamic heterogeneity found in experiments, and reduce the negative correlations that cause excess *pe* fluctuations in MD simulations. We anticipate that adding heterogeneity to MD simulations via a mechanism that decorrelates neighboring regions will also improve agreement with measured response, as has been found for MC simulations [28-33]. Furthermore, these MC simulations mimic the statistics of indistinguishable particles within each region, allowing simple classical models to crudely exhibit some aspects of quantum-like behavior.

### G) Adding quantum-like behavior to MD simulations

We know of no unambiguous experimental evidence for the excess *pe* fluctuations that we find in MD simulations. Insight comes from the excess specific heat measured in most materials at low $T$ [50]. For example, high-purity single crystals at $T$<1 K exhibit excess specific heat that diverges from Debye's theory inversely proportional to a power of $T$ [51,54], superficially similar to the $1/T$ divergence of excess *pe* fluctuations in MD simulations shown in Fig. 3 (A). Furthermore, measured specific heats show slow and non-exponential relaxation [51-52,54], superficially similar to the *ke* relaxation in Fig. 4 (D). Moreover, amorphous materials deviate from Debye's theory at much higher $T$ [52,53], as expected when disorder pushes many 1[st]-neighbor atoms from their robustly harmonic *pe* minima. Excess specific heat at low $T$ is usually attributed to spatially-localized tunneling 2-level systems, but often involving an unknown source, especially in high-purity single crystals. Our MD simulations also exhibit *pe* localization, even in ideal crystals. We suggest that excess *pe* fluctuations in MD simulations involve the Anderson mechanism for localization that requires a broad distribution of energies, as shown in Fig. 6. We speculate that measured excess specific heat at low $T$ may come from phonons that are localized by uncorrelated motion between atoms that are not at the bottom of robustly harmonic wells, even in high-purity single crystals when 2[nd]-neighbor interactions are included. However, we reiterate that the simulations show excess *pe* fluctuations, not excess specific heat. Furthermore, Debye's theory is based on quantum behavior that is absent from our MD simulations.

Debye's theory of specific heat utilizes Bose-Einstein statistics to reduce the occupation of high-frequency phonons, yielding the $T^3$ dependence shown by most measurements at low $T$. Such quantum statistics in realistic materials will preferentially suppress zone-boundary modes, reducing the unrealistic excess *pe* fluctuations that we find in classical MD simulations. Although quantum effects are often used to refine the interactions in classical MD simulations, our results indicate the need to use quantum statistics for accurate thermal fluctuations, at least up to the melting point if long-range interactions are involved (e.g. up to 244 K in NM) [104,105]. Another aspect of quantum behavior could be to facilitate the nanometer-scale heterogeneity that may be necessary to yield consistent thermal dynamics in MD simulations, and to agree with the heterogeneity found in many measurements [52-69]. Alternatively, adding thermodynamic heterogeneity to classical MD simulations could provide a simplified approach to mimic quantum-like statistics and improve agreement with measured response, as has been done with MC simulations [28-33].

## VI. Conclusions

We extend the original goal of Enrico Fermi and his team to compare equilibrium MD simulations based on Newton's laws with standard statistical mechanics based on Boltzmann's factor. We find that the average energies give excellent agreement with the equipartition theorem, but energy fluctuations deviate significantly from the analogous energy fluctuation relation, especially at low temperatures. In fact, excess *pe* fluctuations tend to diverge inversely proportional to *T*, as shown in Fig. 3 (A). Empirically, at the lowest *T* studied, equilibrium *pe* fluctuations can be characterized by an effective local temperature that is an order of magnitude hotter than the temperature for *ke* fluctuations. These deviations occur for fluctuations of individual atoms, as well as for blocks containing 8-6912 atoms inside much larger simulations that are in the thermodynamic limit, using either pure Newtonian dynamics or a Nosé-Hoover thermostat. Figure 2 establishes that abruptly when the interaction cutoff radius is reduced to include only the robustly harmonic interaction between $1^{st}$-neighbor atoms, the *pe* fluctuations are reduced to below the *ke* fluctuations. Thus, excess *pe* fluctuations require interactions that include at least two types of neighbors.

We elucidate the source of anomalous fluctuations using energy correlations as a function of time and distance. Figures 2 and 3 show a clear connection between the excess *pe* fluctuations in small blocks and negative correlations in the *pe* between neighboring blocks, whereas for reduced *pe* fluctuations this correlation is positive. Figures 3 and 4 confirm that anomalous *pe* fluctuations in small blocks involve energy that is localized to a surrounding shell of similarly small blocks. Figure 4 (C) also shows that Δ*pe* oscillates rapidly, and is strongly dissipated before the speed of sound carries excess *pe* much beyond the neighboring blocks, isolating the local dynamics from the large heat reservoir needed to justify Boltzmann's factor. Combining the *pe* of each block with the *pe* of its neighboring blocks yields a modified energy for Boltzmann's factor, or equivalently an effective local temperature that is consistent with the effective temperature deduced from anomalous *pe* fluctuations. Figure 2 shows that excess *pe* fluctuations become reduced *pe* fluctuations abruptly when the interaction cutoff is reduced to yield interactions only between $1^{st}$-neighbor atoms. Figures 2 and 5 show that excess *pe* fluctuations are connected to density fluctuations, but only if there are $2^{nd}$-neighbor interactions. We attribute the mechanism to an incipient local distortion that transiently brings $2^{nd}$-neighbor atoms closer to their $1^{st}$-neighbor distance, which significantly lowers the local interaction energy consistent with the transition to icosahedral structure for small Ar clusters [88], while creating high-energy interfaces that greatly broaden the distribution of energies. Indeed, Fig. 6 shows a fivefold increase in the width of the *pe* distribution when $2^{nd}$-neighbor interactions are added. We suggest that this increased width provides the distribution of local energies necessary for a type of Anderson localization, even in ideal crystals at sufficiently low *T*.

A basic model is presented that quantitatively characterizes various features in the simulations. A crucial ingredient in the model is that Boltzmann's factor must be modified to exclude (or at least reduce) the *pe* from 2nd-neighbor atoms. The mechanism is attributable to the fact that intrinsically anharmonic interactions between 2nd-neighbor atoms do not couple strongly to harmonic normal modes, becoming isolated from the large heat reservoir due to the fast dynamics of zone-boundary modes and weak coupling from anharmonic scattering. In other words, the 2nd-neighbor interactions that dominate the excess *pe* fluctuations are disconnected from Boltzmann's factor so that they should be treated by an adiabatic ensemble, consistent with the localized energy that we find from *pe* correlations. Because most models of practical importance include interactions that are not robustly harmonic, similar excess *pe* fluctuations are expected in most MD simulations.

Our study reveals a fundamental inconsistency in the dynamics deduced from classical MD simulations, theory, and experiments. Specifically, we have shown that excess *pe* fluctuations from MD simulations based on Newton's laws often differ significantly from standard statistical mechanics based on Boltzmann's factor; and that neither approach fully explains the thermodynamic heterogeneity and excess specific heat found by several experimental techniques in disordered materials, and also in high-purity single crystals at sufficiently low temperatures [50-69]. Our MD simulations and analysis show that anomalous *pe* fluctuations in nanometer-sized subsystems come from temporal memory effects and spatial correlations in the surrounding shell of similarly small systems, which are effectively isolated from the large heat reservoir needed for Boltzmann's factor. Prior progress in the theoretical understanding of small systems has been made by adding thermodynamic heterogeneity to statistical mechanics using nanothermodynamics as a guide [15], yielding a nonlinear correction to Boltzmann's factor that facilitates the self-consistent treatment of small systems in thermal contact with an ensemble of similarly small systems [16]. We anticipate that adding thermodynamic heterogeneity to models and algorithms in MD simulations will decorrelate the dynamics between local region in the simulations, allowing access to lower-energy states while reducing the *pe* fluctuations. The resulting nanometer-scale correlations will yield a fundamental size for the crossover from reversible dynamics to irreversible thermodynamics, and for intrinsic coarse-graining in nature. This thermodynamic heterogeneity should also improve agreement between MD simulations, theoretical fluctuation relations, and measured response.

**VII. Acknowledgments**

We are grateful for helpful discussions with S. Abe, O. Beckstein, B. F. Davis, J. Dyre, K. Ghosh, A. Heuer, M. Heyden, Y. Li, N. Newman, J. B. Page, B. M. Rice, S. L. Seyler, and G. H. Wolf. RVC has the pleasure to thank Betsy Rice and her team for their hospitality and assistance during a stay at the Army Research Laboratory. Most of the simulations were performed using the facilities of ASU Research Computing.

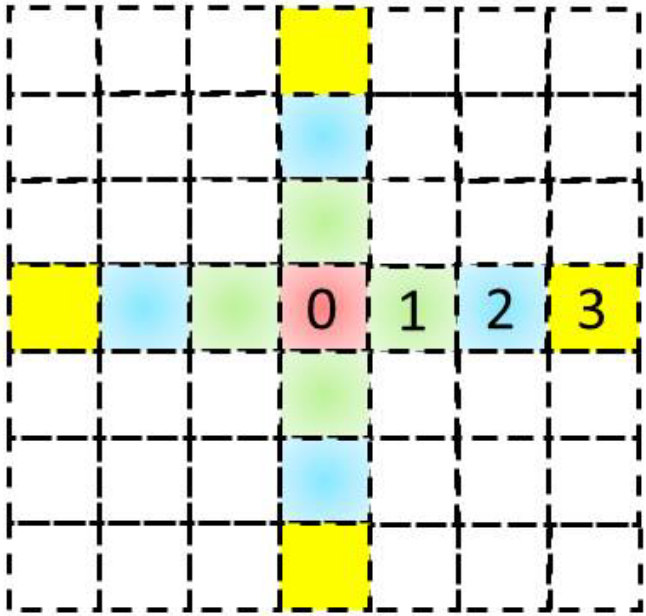

FIG. 1. (Color online) Schematic representation of small blocks that form local systems inside a much larger simulation volume. Each block encompasses an integer number of unit cells, so that they are cube-shaped for the L-J model and orthorhombic for NM. A central block is labeled "0", a $1^{st}$-neighbor block is labeled "1", etc. Periodic boundary conditions are used on all outside surfaces, so that optimal statistics is achieved using ensemble-averaging with each block as the center "0," and time-averaging over the entire production run for equilibrium behavior. Similarly, time-dependent properties are deduced from ensemble averaging the behavior of individual blocks at times separated by $t$, averaged over the entire production run. Each block mimics the textbook example of a system that is in the grand-canonical ensemble. Canonical-ensemble behavior is extracted by selecting the subset of blocks that have the equilibrium number of atoms in each block.

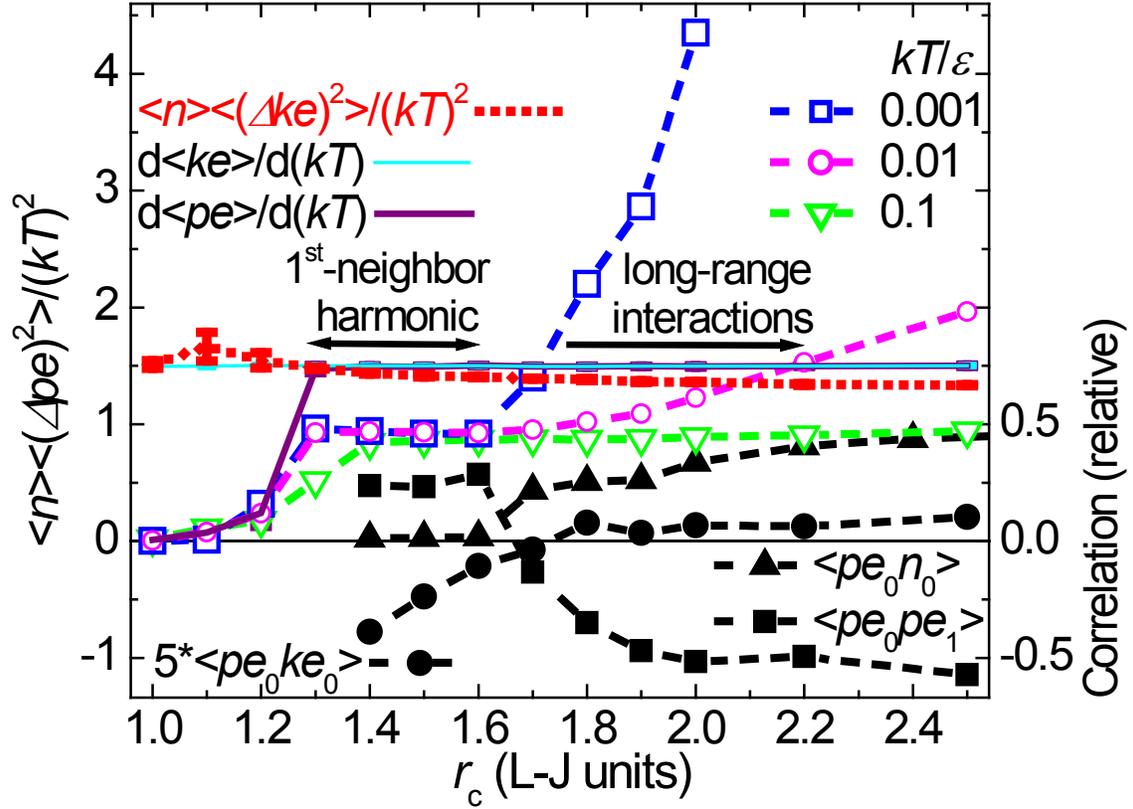

FIG. 2. (Color online) Interaction cutoff radius dependence of normalized energy fluctuations and specific heat (left scale), and correlations (right scale), from blocks having $n=32$ in the L-J model. Note that for $1.3 \leq r_c \leq 1.6$ at low $T$ the primary interaction is effectively harmonic between 1st-neighbor atoms, with significant 2nd-neighbor interactions only if $r_c > 1.6$. Open symbols show fluctuations in $pe$ at three $T$ (given in the legend), with dashed lines connecting the symbols as a guide for the eye. The dotted red line shows analogous fluctuations in $ke$, with error bars from averaging over six $T$ ($kT/\varepsilon=0.0005$ to $0.02$). Solid lines show the $T$-dependent derivatives of $\langle ke \rangle$ (cyan) and $\langle pe \rangle$ (purple), averaged over the same range of $T$, with error bars visible due to the line thickness. Solid symbols (from simulations at $kT/\varepsilon=0.001$) show normalized correlations of $pe$ in each central block with $ke$ (circles), with $n$ (triangles), as well as with $pe$ in neighboring blocks (squares). The EFR predicts identical values for the dotted red and solid cyan lines, as well as for the open symbols and solid purple line, found only for short-range purely repulsive forces, $r_c<1.3$. Note that in general, positive $\langle pe_0 pe_1 \rangle$ yields $pe$ fluctuations that are significantly below the EFR, whereas negative $\langle pe_0 pe_1 \rangle$ yields excess $pe$ fluctuations that diverge with decreasing $T$.

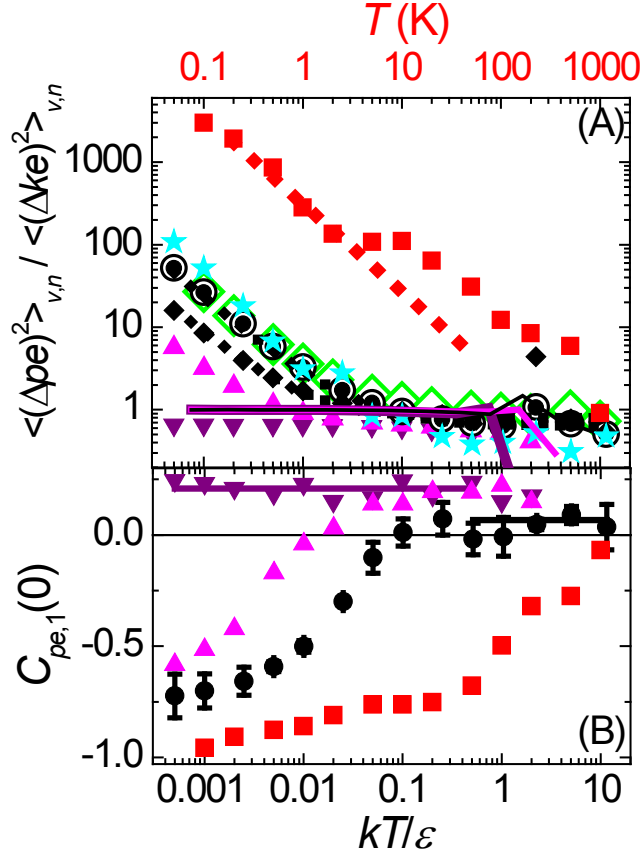

FIG. 3 (Color online) (A) Symbols show the ratio of local energy fluctuations, from simulations of NM (squares) with $T$ in K (top scale), and the L-J model (other symbols) with $T$ in L-J units (bottom scale). These $T$ were calculated from the kinetic energy per atom, averaged over all blocks that have the correct $<n>$. Open circles are from blocks containing $<n>=32$ atoms inside simulations containing $N=55,296$ atoms, solid symbols are from blocks with $<n>=32$ (circles) and $<n>=2048$ (diamonds) with $N=442,368$, while stars come from individual atoms. Open diamonds show simulations using a Nosé-Hoover thermostat, while all other simulations use pure Newtonian dynamics. Dotted lines show the $1/T$ divergence of excess $pe$ fluctuations as $T\rightarrow 0$. Triangles show analogous simulations to the solid circles, except that the interaction range is reduced to include only 1st-neighbor atoms $r_c=1.5$ (down), or also 2nd-neighbor atoms $r_c=2.0$ (up). Solid lines that vary from thick to thin show the ratio of average specific heats, $\frac{d<pe>/d(kT)}{d<ke>/d(kT)}$, for interaction cutoff radii in the L-J model of $r_c=1.5$, 2.0, and 6.0, with peaks that identify the melting temperatures of $kT/\varepsilon=0.75$, 1.5, and 1.5, respectively. (B) Initial $pe$ correlations between 1st-neighbor blocks in the L-J model with $r_c=1.5$ (down triangles), 2.0 (up triangles), and 6.0 (circles), and in NM (squares). Solid lines show average values where the behavior is independent of $T$. Down triangles in (A) and (B) show that $r_c=1.5$ yields fluctuations below the EFR prediction, with positive initial correlations, opposite to the low-$T$ behavior of analogous simulations with long-range interactions (other symbols).

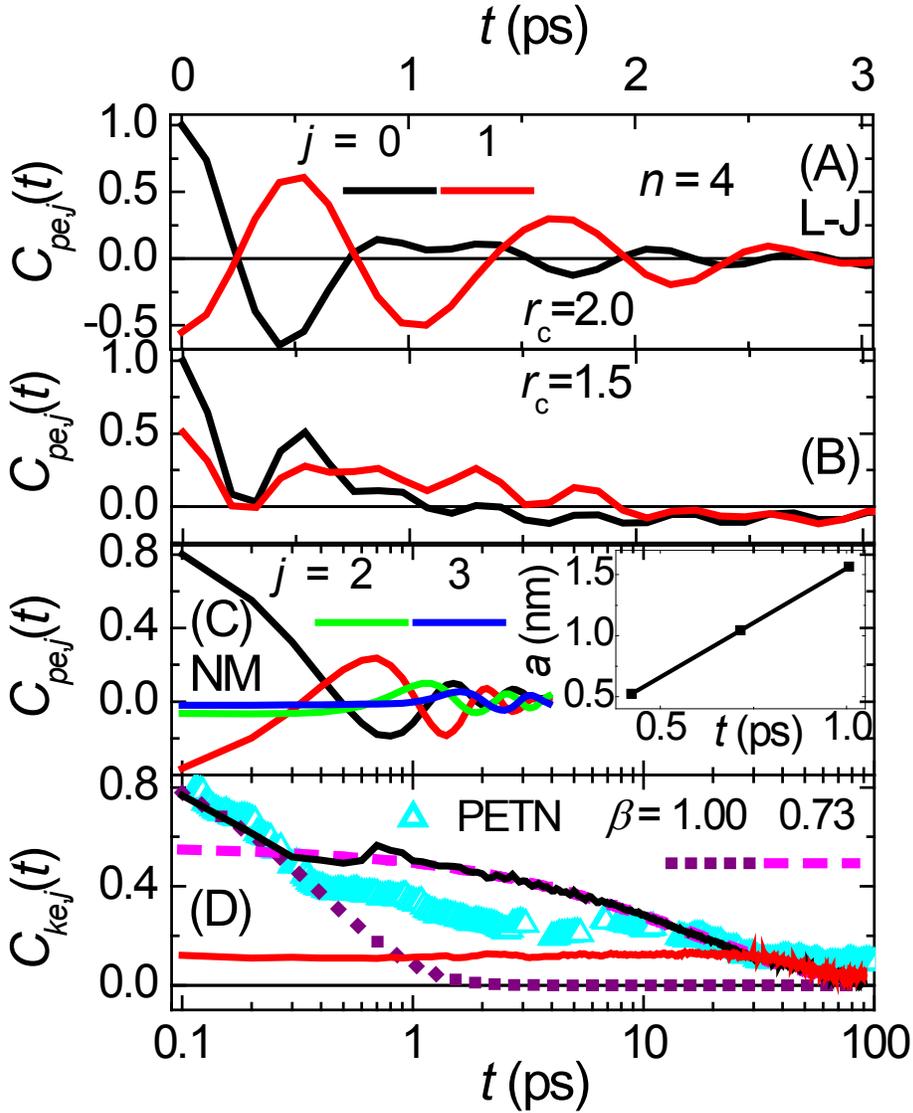

FIG. 4. (Color online) Upper two panels show time-dependent *pe* autocorrelations (black) and correlations between 1st-neighbor blocks (red), with time on a linear scale (upper axis) using 2.156 ps/(L-J unit). These upper two panels are from the L-J model at $kT/\varepsilon=0.0005$ in blocks having $n=4$ with cutoff radii of 2.0 (A) and 1.5 (B). Lower two panels show correlations in *pe* (C) and *ke* (D), from the *a*-axis of NM at $T=100$ K, with time on a logarithmic scale (lower axis). (C) also shows *pe* correlations to more-distant blocks: $j=2$ (green) and $j=3$ (blue). The inset of (C) shows the distance between 1st-, 2nd-, and 3rd-neighbor blocks as a function of time at which the *pe* correlations reach their first maximum, which yields the solid line of slope 1.11 nm/ps (1110 m/s). Broken lines in (D) are from fits to the *ke* autocorrelation showing exponential relaxation at short times (dotted) and stretched-exponential at long times (dashed). Triangles in (D) show temperature as a function of time from measurements of PETN [85], with the *x*- and *y*-axes offset and/or multiplied by a factor to put them on the same scale as the simulations.

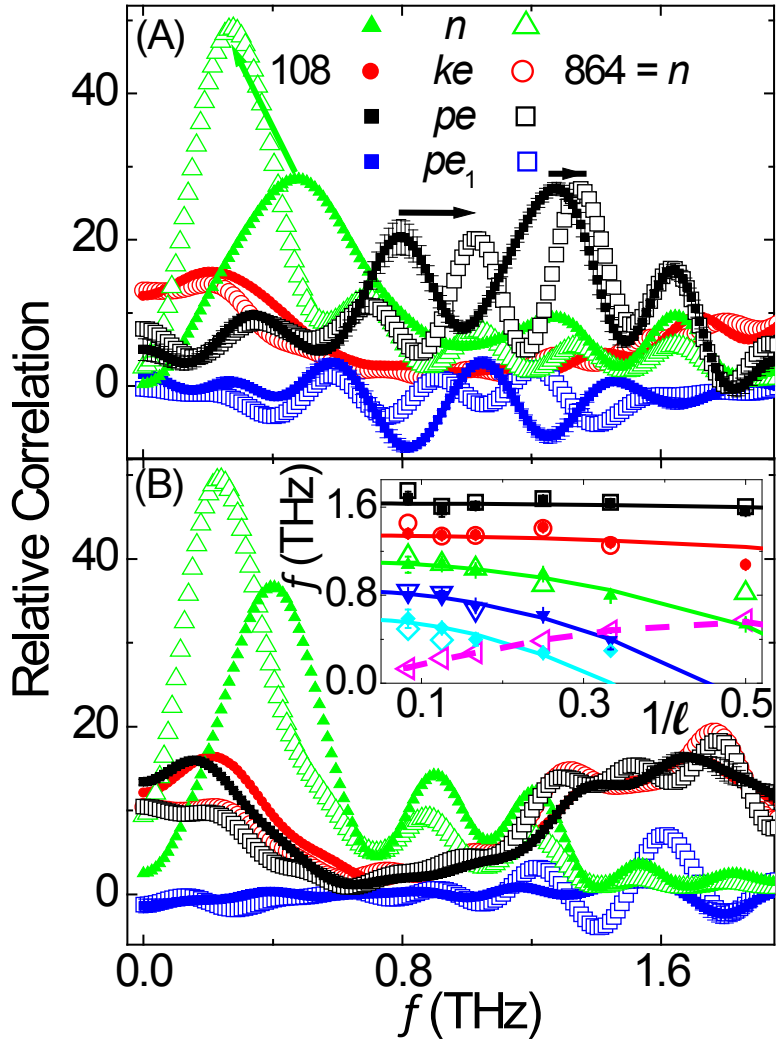

FIG. 5. (Color online) Frequency-dependent correlations in the L-J model from simulations with interaction cutoff radii of $r_c$=2.0 (A) and $r_c$=1.5 (B). Symbols come from the Fourier transform of the normalized autocorrelations in $n$ (green triangles), $ke$ (red circles), and $pe$ (black squares). Blue squares show correlations in $pe$ between 1$^{st}$-neighbor blocks. Each spectrum comes from averaging behavior from all blocks having a constant number of atoms $n$=108 (solid symbols) or $n$=864 (open symbols), except for fluctuations in $n$ which were obtained by shifting the block positions to align with the equilibrium atomic planes so that $n$ fluctuates as atoms move. Each spectrum also involves averaging from three temperatures ($kT/\varepsilon$ = 0.0005, 0.001, and 0.002), with error bars for the solid squares (visible when larger than the symbol size). Arrows in (A) indicate how three normal modes (peaks) shift when the block size changes. Only fluctuations in $n$ show these distinct peaks in the effectively harmonic lattice, (B). Symbols in the inset of (B) show the $1/\ell$ dependence of the peaks from (A) for the $pe$ correlations (solid) and $n$ correlations (open), with lines from fits to the data using a sine function (dashed) or quadratic function (solid), see text.

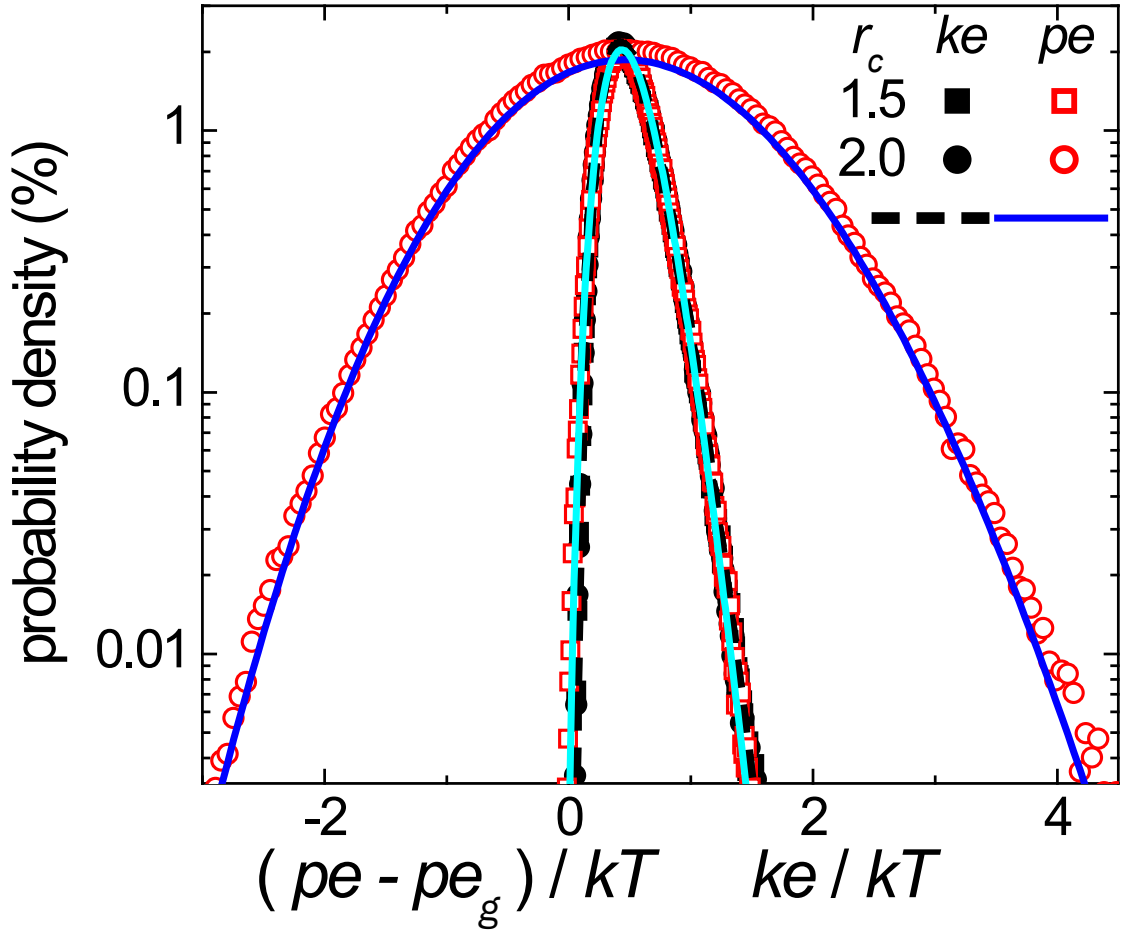

FIG. 6. (Color online) Energy distributions as a function of reduced energy. Simulations are of the L-J model at $kT/\varepsilon = 0.001$ with cut-off radii of $r_c$=1.5 (squares) and 2.0 (circles) from blocks containing a single unit cell ($n = 4.00$). Open symbols show $pe$ probabilities as a function of $(pe-pe_g)/kT$, where $pe_g$ is the ground state energy as $T\rightarrow 0$. Solid symbols (often hidden by open squares) show $ke$ probabilities as a function of $ke/kT$. Although not shown here, dividing by $kT$ effectively removes all $T$ dependence of the $ke$ distribution, and of the $pe$ distribution for $r_c$<1.6, whereas the width of the $pe$ distribution increases sharply with decreasing $T$ when $r_c$>1.6. The dashed line shows the fit to the $ke$ probabilities using the canonical-ensemble density of states $(AE/kT)^\gamma e^{-BE/kT}$, yielding $\gamma=4.99\pm0.06$, consistent with $<n>D/2-1=5.00$ expected for a finite system of ideal-gas particles. Solid lines show fits to the $pe$ distribution, yielding $\gamma=7.5\pm0.2$ for $r_c$=1.5 (cyan) and 325±1 for $r_c$=2.0 (blue).